\documentclass[twocolumn,superscriptaddress,aps,longbibliography,showpacs,preprintnumbers,amsmath,amssymb,floatfix]{revtex4-1}
\usepackage{titlesec}
\usepackage{amsmath}
\usepackage{graphicx}
\usepackage[ansinew]{inputenc}
\usepackage{array}
\usepackage{color}
\usepackage{amsxtra}
\usepackage{amstext}
\usepackage{amssymb}
\usepackage{latexsym}
\usepackage{gensymb}
\usepackage{dsfont}
\usepackage{braket}
\usepackage[caption=false]{subfig}
\usepackage[makeroom]{cancel}

\usepackage{dcolumn}
\usepackage{bm}
\usepackage{bbm}
\usepackage{braket}
\usepackage{mathrsfs}
\usepackage{amstext}
\usepackage{euscript}
\usepackage{txfonts}
\usepackage{relsize}

\usepackage[unicode=true,
 bookmarks=false,
 breaklinks=false,pdfborder={0 0 1},backref=false,colorlinks=true,citecolor=blue]
 {hyperref}

\makeatletter
\@ifundefined{textcolor}{}
{%
 \definecolor{BLACK}{gray}{0}
 \definecolor{WHITE}{gray}{1}
 \definecolor{RED}{rgb}{1,0,0}
 \definecolor{GREEN}{rgb}{0,1,0}
 \definecolor{BLUE}{rgb}{0,0,1}
 \definecolor{CYAN}{cmyk}{1,0,0,0}
 \definecolor{MAGENTA}{cmyk}{0,1,0,0}
 \definecolor{YELLOW}{cmyk}{0,0,1,0}
}

\makeatletter
\renewcommand*\env@matrix[1][*\c@MaxMatrixCols c]{%
  \hskip -\arraycolsep
  \let\@ifnextchar\new@ifnextchar
  \array{#1}}
\makeatother

\newcommand{\cref}[1]{Ref.\,\cite{#1}}

\makeatother

\begin{document}


\title{Evaluating Single-mode Nonclassicality}
\date{\today}
\author{Wenchao Ge}
\affiliation{Institute for Quantum Science and Engineering (IQSE) and Department of Physics and Astronomy, Texas A\&M University, College Station, TX 77843-4242, USA}
\author{M. Suhail Zubairy}\affiliation{Institute for Quantum Science and Engineering (IQSE) and Department of Physics and Astronomy, Texas A\&M University, College Station, TX 77843-4242, USA}

\begin{abstract}
Quantifying nonclassicality of a bosonic mode is an important but challenge task in quantum optics. Recently, the first nonclassicality measure based on the concept of operational resource theory has been proposed [Phys. Rev. Research \textbf{2}, 023400 (2020)], which shows several crucial properties as a resource for quantum metrology. Here we apply the measure to evaluate and categorize different classes of nonclassical states. We discover a class of states that can achieve the maximum nonclassicality in the asymptotic limit of large mean number of excitations. These states can provide the same sensitivity in place of a squeezed vacuum in a sensing scheme, e.g., the LIGO experiments. We also discover that the nonclassicality of certain states can be greatly improved with a single-photon addition. Besides, we explore some examples on how to evaluate analytically the nonclassicality of mixed states, which are in general difficult to calculate due to the convex roof nature of the measure. Our results will be useful for preparing and utilizing nonclassical states for  applications in precision sensing tasks, such as quadrature sensing and phase sensing in a Mach-Zehnder interferometer.
\end{abstract}

\maketitle

\section{Introduction}
Nonclassical states of bosonic modes, such as optical fields \cite{SZ}, motional states of trapped ions \cite{PhysRevLett.76.1796} or mechanical oscillators in optomechanical systems \cite{PhysRevA.56.4175}, are important resources for quantum-enhanced technologies \cite{loncar2019development}, including quantum communication \cite{Braunstein98,Milburn99,PhysRevLett.88.057902}, quantum computation \cite{PhysRevA.59.2631,Lloyd99,Bennett:2000aa,Knill:2001aa, PhysRevLett.119.030502}, and quantum metrology \cite{Caves81,Holland:1993aa,Braunstein:1994aa, Pezze08,Xu:2012aa,PhysRevLett.110.163604, Aasi:2013aa,Lang13, Liu:2013aa, TanPRA14,toth2014quantum, Demkowicz-Dobrzanski:2015aa, GePRL2018, Mccormick2018,PhysRevLett.124.171101,PhysRevLett.124.171102,PhysRevApplied.13.024037,polino2020photonic}. This is enabled by the superposition of coherent states, the most intriguing feature of nonclassical states. Due to its potential as an important resource, quantitative understanding of single-mode nonclassicality is crucial to the field of quantum optics and related subjects. 

The minimum requirement for obtaining a nonclassicality measure is that a quantification is nonnegative for any state and it is zero only if the state is classical. Based on this requirement, there have been many different measures proposed for quantifying single-mode nonclassicality, including the nonclassical distance \cite{Marian:02, hillery1987nonclassical}, the nonclassicality depth \cite{Lee:91}, the entanglement potential \cite{AsbothPRL05}, and quantifications via the Schmidt rank \cite{GehrkePRA12, Vogel:14}. Many of these interesting definitions capture some aspects of the intriguing nonclassical feature. For example, the quantifications using the Schmidt rank are defined to be proportional to the minimum number of coherent superpositions in a quantum state \cite{GehrkePRA12}.

Recently, quantifying nonclassicality has been studied on a stricter notion based on resource theories (RTs) \cite{Tan17, Streltsov17, RevModPhys.91.025001}. According to RTs, all quantum states can be categorized into two groups, one being free states and the other being resource states. In addition, RTs define a set of operations as free operations such that they can not increase the quantity of interest on any state. For the resource theory of nonclassicality \cite{Tan17}, the free states are classical states and the resource states are nonclassical states. A natural choice of free quantum operations for nonclassicality is the set of classical operations, such as applying phase shifts and beam splitters. By being a stricter definition, an RT nonclassicality measure has to both satisfy aformentioned non-negativity and be monotonically non-increasing under any classical operations \cite{Streltsov17}.

While RTs provide the basic methodology for defining a measure of nonclassical states in terms of resources \cite{RevModPhys.91.025001}, they do not necessarily require a measure to quantify the ability of quantum states to provide enhanced performance for certain tasks, such as precision sensing, which is referred as "operational" \footnote{This definition of operational is different from the usual definition, where it means the ability to be converted and manipulated \cite{PhysRevResearch.2.012035, Winter16}.}. Recently, there have been some efforts devoted to the study of an operational resource theory (ORT) of nonclassicality \cite{YadinPRX18, Kwon19, ge2019operational}. An important RT measure of nonclassicality in terms of mean quadrature variance has been proposed by Yadin \emph{et al.}~\cite{YadinPRX18}  and Kwon \emph{et al.}~\cite{Kwon19} independently. The measure has the meaning of the metrological enhancement beyond the standard quantum limit for an averaged sensing task for pure states, however, it is unknown if it has a direct operational meaning in terms of metrology for mixed states \cite{Kwon19}. Ge \emph{et al.} \cite {ge2019operational} proposed the first operational resource theory measure of nonclassicality, which satisfies the minimal requirements of an RT, quantifies the ability to perform quadrature sensing for pure states, and is a tight upper bound for the latter for mixed states. Interestingly, this measure also quantifies macroscopicity \cite{Frowis2018} of nonclassical states in terms of the averaged size of coherent superpositions. 

In this work, we apply the ORT measure to evaluate the nonclassicality of single-mode quantum states, both pure and mixed. While some pure-state examples have been given in Ref. \cite{ge2019operational} to show the concept and the crucial properties of ORT of nonclassicality, there are some interesting questions remaining about evaluating single-mode nonclassicality using the measure. First, the ORT measure suggests the maximum nonclassical state for a fixed energy is a squeezed vacuum state \cite{ge2019operational}. Then an interesting question is that is it possible to find other states that may achieve this maximum nonclassicality asymptotically in certain limiting case. Second, are there any simple operations to greatly enhance nonclassicality of a quantum state? Third, the ORT measure for a mixed state is based on a convex roof construction, where a minimization is over all possible decomposition of a quantum state. Then an important question is how to calculate nonclassicality for mixed states at least for some classes of states.

We answer these questions in this work by evaluating many examples of single-mode quantum states, including Fock states, squeezed coherent states, cat states, single-photon added states, and mixed states in diagonal Fock basis. We categorize the sets of quantum states using the ORT measure and its relation to quadrature sensing. In particular, we categorize the group of nonclassical pure states into three classes depending on their nonclassicality. We find that there is a class of states that are as nonclassical as a squeezed vacuum in the asymptotic limit of a large number of average excitations, which provides interesting alternatives for quantum metrology. We also investigate the nonclassicalities of a quantum state before and after single-photon addition. Our results show that single-photon operations can greatly increase the nonclassicality of a state quantified by the ORT measure, which will be important for preparing strong nonclassical states using weaker nonclassical states. For mixed states with finite dimensions, we show some examples of calculating the nonclassicality by analytically finding the convex roof using the measure. Our results show that the nonclassicality of some Fock state superpositions may not be affected by coupling to an environment, e.g., phase damping. For mixed states with infinite dimensions, we calculate lower bounds for the measure. 

The paper is organized as follows. In Sec. \ref{sec2}, we introduce the ORT measure of nonclassicality and its relation to two important quantum sensing tasks. In Sec. \ref{sec3}, we investigate extensively examples of nonclassical states using the ORT measure. In Sec. \ref{sec4}, we compare the ORT measure with some existing measures of nonclassicality. We summarize the main results of this work in Sec. \ref{sec5}. 

\section{Some Basics of Nonclassical States \label{sec2}}

\subsection{Operational Nonclassicality Measure}

We begin by introducing the definition of a nonclassical state. A single-mode quantum state $\hat{\rho}$ can be represented using the Glauber-Sudarshan $P$ function \cite{Glauber63, Sudarshan63} as 
\begin{eqnarray}
\hat{\rho}=\int P(\alpha,\alpha^{\ast})\ket{\alpha}\bra{\alpha}d^2\alpha.
\label{eq:P-function}
\end{eqnarray}
The state $\hat\rho$ is defined as classical if the probability distribution function $P(\alpha,\alpha^{\ast})$ is positive definite mimicking a classical probability density over the coherent states $|\alpha\rangle$. The state is nonclassical if $P(\alpha,\alpha^{\ast})$ is singular and not positive definite \cite{Lee:91,SZ}.

Now we introduce the operational resource theory measure of nonclassicality given by \cite{ge2019operational}
\begin{align}
\mathcal{N}\left(\hat{\rho}\right) & = \min_{\{p_j,\ket{\psi_j}\}}\biggl\{\max_{\mu}\sum_j p_j\langle \psi_j | (\Delta\hat{X}_{\mu})^2 | \psi_j \rangle \biggr\}-\frac{1}{2}\nonumber\\
&=\min_{\{p_j,\ket{\psi_j}\}} \biggl\{\sum_j p_j\left(\bar{n}_j-|\bar{\alpha}_j|^2\right)+\biggl|\sum_j p_j\left(\bar{\xi}_j-\bar{\alpha}_j^2\right)\biggr|\biggr\} ,  \label{eq:n-mixed}
\end{align}
where the minimization is over all possible ensembles with $\hat{\rho}=\sum_jp_j \ket{\psi_j}\bra{\psi_j}$ $\big(p_j>0$ and $\sum_jp_j=1\big)$ and the maximization is over all possible quadratures, defined by $\hat{X}_{\mu}=i\left(e^{-i\mu}\hat{a}^{\dagger}-e^{i\mu}\hat{a}\right)/\sqrt{2}$ in which $\hat{a}$ is the annihilation operator for the bosonic mode and $\mu\in [0,2\pi]$.  In the second line of Eq. \eqref{eq:n-mixed}, we have used the moments $\bar{n}_j \equiv \bra{\psi_j}\hat{a}^{\dagger}\hat{a}\ket{\psi_j}$,  $\bar{\xi}_j \equiv \bra{\psi_j}\hat{a}^2\ket{\psi_j}$,  and $\bar{\alpha}_j \equiv \bra{\psi_j}\hat{a}\ket{\psi_j}$. Hence, one can see that $\mathcal{N}\left(\hat{\rho}\right)=0$ for a classical state using the coherent-state decomposition from the definition in Eq. \eqref{eq:P-function}.

Unlike the nonclassicality witness via squeezing (minimum variance) \cite{SZ}, the definition Eq. \eqref{eq:n-mixed} manifests itself as the maximum quadrature variance, which is shown to give rise to the relation to quantum-enhanced metrology.
For pure states, the measure reduces to
\begin{align}
    \mathcal{N}(|\psi\rangle) = \bar{n}-|\bar{\alpha}|^2+\left|\bar{\xi}-\bar{\alpha}^2\right|,
    \label{Npure}
\end{align} 
where $\bar{n} \equiv \langle \hat{a}^{\dagger}\hat{a}\rangle$,  $\bar{\xi}\equiv\langle \hat{a}^2\rangle$,  and $\bar{\alpha}\equiv\langle \hat{a}\rangle$. 

In general, the definition $\mathcal{N}\left(\hat{\rho}\right)$ has been shown \cite{ge2019operational} to satisfy (i) Non-negativity: $\mathcal{N}\left(\hat{\rho}\right)\ge0$ for any state $\hat{\rho}$ where the equality holds if and only if $\hat{\rho}$ is classical; (ii) Weak monotonicity: $\mathcal{N}$ cannot be increased by any classical operation $\Lambda$, i.e., $\mathcal{N}\left(\Lambda[\hat{\rho}]\right) \le \mathcal{N}\left(\hat{\rho}\right)$; (iii) Convexity: $\sum_jp_j\mathcal{N}\left(\hat{\rho}_j\right)\ge\mathcal{N}\left(\sum_jp_j\hat{\rho}_j\right)$ for any quantum states $\hat{\rho}_j$ and probabilities $p_j$. A classical operation can be augmentation by any number of classical states, the application of any passive linear optical operations and displacements, or tracing out of the auxiliary modes. The first two conditions are the minimum requirements for a meaningful measure in the RT of nonclassicality. 

The concept of the ORT of nonclassicality \cite{YadinPRX18,Kwon19,ge2019operational} in addition requires a RT measure to have an operational meaning such that it 
 relates to the ability for performing certain tasks. This will be discussed in the following when we introduce the tasks of quadrature sensing and phase sensing in a Mach-Zehnder interferometer.

\subsection{Quantum Metrology with Nonclassical States}
 \begin{figure}[t]
\leavevmode\includegraphics[width = 1\columnwidth]{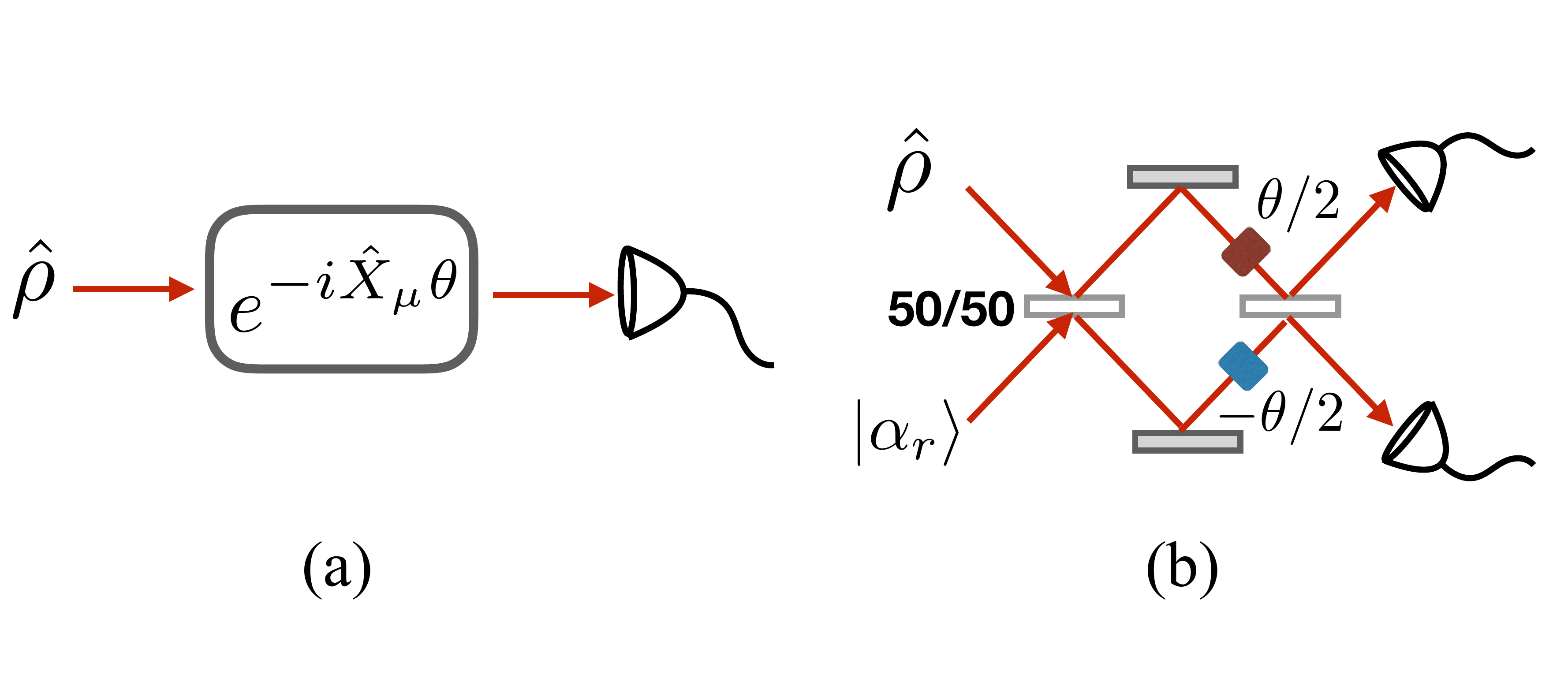}
\caption{Quantum sensing tasks. (a) Quadrature sensing using a single-mode quantum state $\hat\rho$. (b) Phase sensing in a balanced Mach-Zehnder interferometer using $\hat\rho$ with a classical resource, i.e., a coherent state $\ket{\alpha_r}$.} 
\label{fig:sensing} 
\end{figure} 
\subsubsection{Quadrature Sensing}
In quadrature sensing, the task is to estimate an unknown parameter $\theta$ that is encoded in a state $\hat\rho$ via unitary dynamics, i.e., $\hat{\rho}(\theta)=e^{-i\hat{X_{\mu}}\theta}\hat\rho e^{i\hat{X_{\mu}}\theta}$ as shown in Fig.~\ref{fig:sensing}~(a). Quadrature sensing has many applications, such as continuous-variable quantum crytography\cite{PhysRevLett.88.057902}  and mechanical displacement sensing \cite{Hoff:2013aa}. According to the quantum Cramer-Rao bound \cite{Braunstein:1994aa}, the estimation sensitivity of an unbiased estimator $\Theta$ for the parameter $\theta$ satisfies
\begin{align}
\Delta^2\Theta\ge \frac{1}{MF_X\left(\hat{\rho}\right)},
\label{eq:CRB}
\end{align}
where $M$ is the number of repetitions and $F_X\left(\hat{\rho}\right) = 4\max_\mu \Biggl[ \min_{\{p_j,\ket{\psi_j}\}}\biggl\{\sum_j p_j\langle \psi_j | (\Delta\hat{X}_\mu)^2 | \psi_j \rangle \biggr\} \Biggr]$ is the optimized quantum Fisher information (QFI) over $\mu$ for the state $\hat{\rho}(\theta)$ \cite{toth2014quantum, Demkowicz-Dobrzanski:2015aa}. For a classical state, $F_X\left(\hat{\rho}\right) \le2$ and $\Delta\Theta=1/\sqrt{2M}$ is defined as the standard quantum limit (SQL) in quadrature sensing. Therefore, the metrological power for quadrature measurement is $\mathcal{W}\left(\hat{\rho}\right) \equiv 1/4\max[F_X\left(\hat{\rho}\right)-2,0]$ \cite{Kwon19, ge2019operational}, quantifying the amount of metrological advantage beyond the SQL. The operational meaning of the measure $\mathcal{N}$ is given by the following relation with the metrological power $\mathcal{W}(\hat{\rho})$  \cite{ge2019operational}
 \begin{align} 
      \mathcal{N}\left(\hat{\rho}\right) \ge \mathcal{W}(\hat{\rho}),
 \label{eq:ineq1} 
 \end{align}
 where the equality holds when $\hat\rho$ is a pure state, meaning that every nonclassical pure state is useful for beating the SQL in quadrature sensing.
 
Using the concepts of $\mathcal{N}$ and $\mathcal{W}$ and their relation, we can visualize different sets of states in Fig. \ref{fig:state}. For example, the set of classical states is given by $\mathcal{N}\left(\hat\rho_{\text{cl}}\right)=0$ and the set of metrological useful states is given by $\mathcal{W}\left(\hat\rho_{\text{mp}}\right)>0$. In between, there is a region of nonclassical states with a zero metrological power, i.e., $\mathcal{N}\left(\hat\rho_{\text{nc}}\right)>0$ and  $\mathcal{W}\left(\hat\rho_{\text{nc}}\right)=0$, as shown in the figure. Since every nonclassical pure state has a nonzero metrological power, a nonclassical state in this region must be a mixed state, which will be discussed in Sec. \ref{sec3b}.

\subsubsection{Phase Sensing\label{sec2b}}
Now we introduce another metrological task in terms of phase sensing in a balanced Mach-Zehnder interferometer (MZI) as shown in Fig.~\ref{fig:sensing}~(b). The MZI is a paradigmatic model in optical metrology \cite{TanPRA14, Demkowicz-Dobrzanski:2015aa, Caves81,Pezze08} with the benchmark example by feeding a coherent state and a squeezed vacuum state \cite{Caves81,Pezze08,PhysRevLett.124.171101,PhysRevLett.124.171102}, which has been used in the LIGO experiments for quantum enhanced sensitivity \cite{ Aasi:2013aa}. Here this idea is generalized by feeding a coherent state $\ket{\alpha_r}$ and a quantum state $\hat\rho$ \cite{ge2019operational}. For a given input state $\hat\rho$, it has been shown that the optimal precision in estimating the phase difference between two paths in a MZI is achieved by choosing the first beam-splitter to be $50/50$, i.e., balanced \cite{Jarzyna12,Hofmann:2009aa, ge2019operational}. 

The optimal QFI at the MZI is given by $F_{\theta}^{\text{MZI}}\left(\hat{\rho}\right)= N +\frac{\left|\alpha_r\right|^2}{2}\left[F_X\left(\hat{\rho}\right)-2\right]$, where $N=|\alpha_r|^2+\bar{n}$ is the mean number of total input photons. Similar to quadrature sensing, the QFI relates to phase estimation sensitivity in the MZI as $\Delta^2\Theta\ge 1/MF_{\theta}^{\text{MZI}}\left(\hat{\rho}\right)$ \cite{Demkowicz-Dobrzanski:2015aa}. Therefore, two implications can follow from the expression of the QFI \cite{ge2019operational}: \\
(i) $F_{\theta}^{\text{MZI}}\left(\hat{\rho}\right)>N\Leftrightarrow F_X\left(\hat{\rho}\right)>2$, meaning that achieving sensitivity beyond the SQL in quadrature sensing using $\hat\rho$ is equivalent to that in phase sensing in the MZI. \\
(ii) Heisenberg-limited phase sensing, i.e.,  $F_{\theta}^{\text{MZI}}\sim N^2$, can be achieved when $F_X\left(\hat{\rho}\right)-2\sim \bar{n}$. In particular, this condition is met when $\mathcal{N(\hat\rho)}\sim \bar{n}$ for pure states by employing the equality in Eq. \eqref{eq:ineq1}. We define $\mathcal{N}_{\bar{n}}\left(\hat\rho\right)\equiv\mathcal{N}\left(\hat\rho\right)/\bar{n}$ as the nonclassicality per unit energy.  Then the Heisenberg-limited sensing can be achieved when $\mathcal{N}_{\bar{n}}\sim 1$.

\section{Evaluating nonclassicality\label{sec3}}
\begin{figure}[t]
\leavevmode\includegraphics[width = 0.8 \columnwidth]{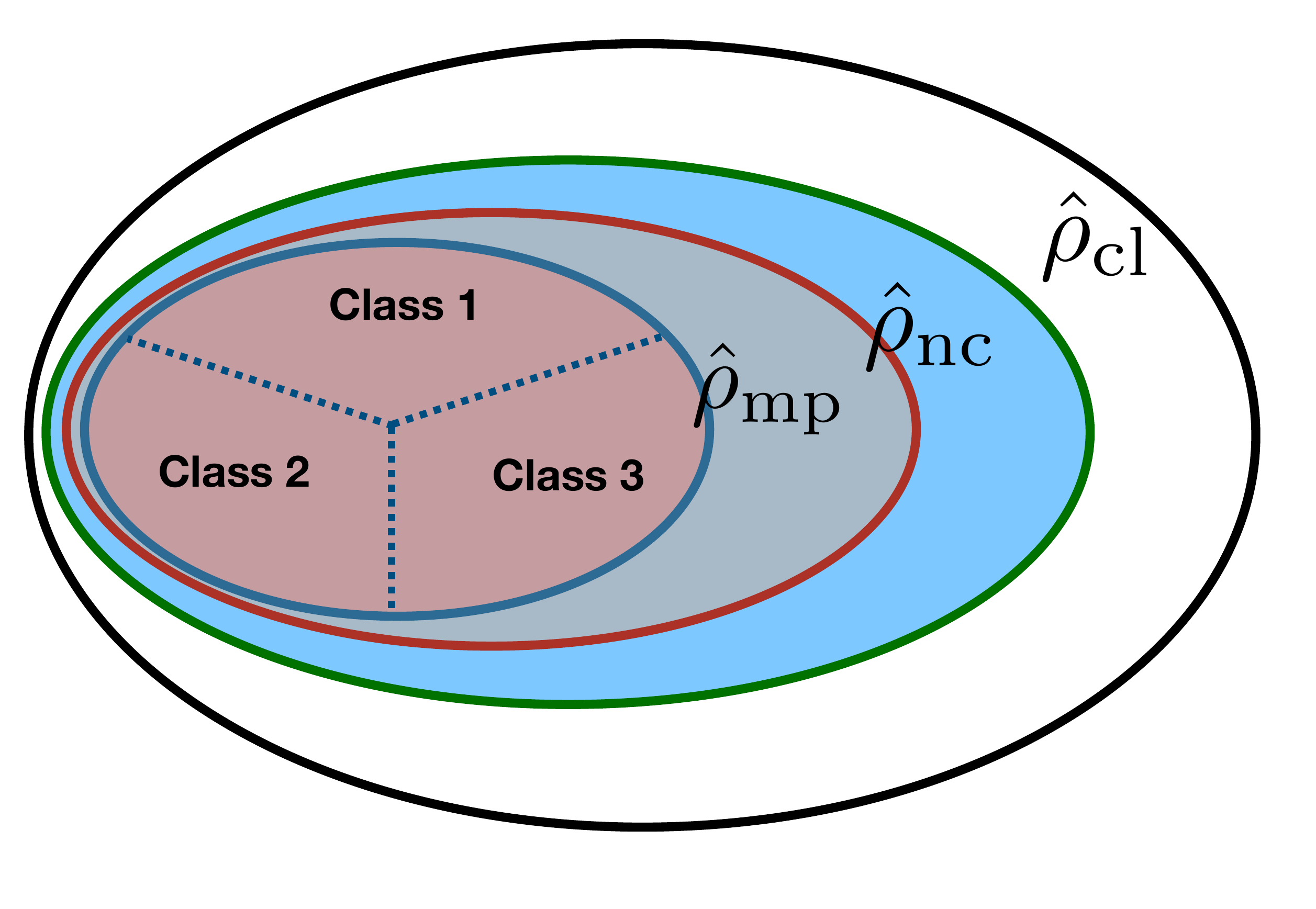}
\caption{Sets of states categorized by the ORT measure $\mathcal{N}$ and the metrological power $\mathcal{W}$. The white region between the black (the largest) and the green (the second largest) ovals corresponds to the set of classical states, i.e., $\mathcal{N}\left(\hat\rho_{cl}\right)=0$. The region inside the green (the second largest) oval is the set of nonclassical states, i.e., $\mathcal{N}\left(\hat\rho_{nc}\right)>0$. The region inside the red (the third largest) oval is the set of nonclassical states with nonzero metrological power, i.e., $\mathcal{W}\left(\hat\rho_{mp}\right)>0$. The region inside the blue (the smallest) oval is the set of nonclassical pure states, which is further categorized into three different classes according to their nonclassicality per unit energy (see the text in Sec.~\ref{sec3a} for details).} 
\label{fig:state} 
\end{figure} 

\subsection{Nonclassicality for pure states\label{sec3a}}
We have shown that nonclassical pure states have the ability to achieve sensitivity beyond the SQL in both displacement sensing and interferometric phase sensing schemes. According to Eq. \eqref{eq:ineq1}, the amount of nonclassicality of a pure state has a one-to-one correspondence to its power for quantum-enhanced metrology, which is the region in the blue oval (the smallest oval) in Fig. \ref{fig:state}. Furthermore, we can categorize the nonclassical pure states into three classes using the nonclassicality per unit energy:\\
\noindent (i)~~~Class 1: $\lim_{\bar{n}\rightarrow \infty}\mathcal{N}_{\bar{n}}=2$;\\
\noindent (ii)~~Class 2: $1\le\lim_{\bar{n}\rightarrow \infty}\mathcal{N}_{\bar{n}}<2$;\\
\noindent (iii)~Class 3: $0<\lim_{\bar{n}\rightarrow \infty}\mathcal{N}_{\bar{n}}<1$.\\
In addition, we study single-photon added states and how they can be put into these categories.

\subsubsection{Class 1: The asymptotic maximum nonclassical states $\lim_{\bar{n}\rightarrow \infty}\mathcal{N}_{\bar{n}}=2$}
It has been shown that a squeezed vacuum state $\ket{\xi}$ has the maximum nonclassical state per unit energy \cite{ge2019operational} with $\mathcal{N}_{\bar{n}}=1+\sqrt{1+1/\bar{n}}$. Therefore, it is the most useful state in the MZI for phase sensing with a coherent input \cite{Lang13, Pezze08}. Here we investigate the class of states that can achieve the asymptotic maximum nonclassicality, i.e., $\lim_{\bar{n}\rightarrow \infty}\mathcal{N}_{\bar{n}}=2$, which could provide alternative possibilities to a squeezed vacuum in quantum-enhanced metrology.

Consider a pure state $\ket{\psi}=\sum_{k=1}^L c_k\ket{\alpha_k}$ that is a superposition of $L$ coherent states $\ket{\alpha_k}$. The complex amplitudes $\alpha_k$ and the coefficients $c_k$ satisfy the normalization condition $\sum_{j,k}c_jc_k^{\ast}f_{jk}=1$, where $f_{jk}=\braket{\alpha_k|\alpha_j}$. Using this representation, we find a class of cat states that can achieve the asymptotic value of maximum nonclassicality $2\bar{n}$ in the limit of $\bar{n}\gg1$. One way to obtain some of these states is to choose $\bar{\alpha}=0$ and $\bar{n}=|\bar{\xi}|$. Using the superposition of coherent states, we find
\begin{align}
\bar{\alpha}=\bra{\psi}\hat{a}\ket{\psi}&=\sum c_jc_k^{\ast}\alpha_jf_{jk}\approx\sum |c_j|^2\alpha_j, \nonumber\\
\bar{\xi}=\bra{\psi}\hat{a}\hat{a}\ket{\psi}&=\sum c_jc_k^{\ast}\alpha^2_jf_{jk}\approx\sum |c_j|^2\alpha^2_j, \nonumber\\
\bar{n}=\bra{\psi}\hat{a}^{\dagger}\hat{a}\ket{\psi}&=\sum c_jc_k^{\ast}\alpha_j\alpha_k^{\ast}f_{jk}\approx\sum |c_j|^2\left|\alpha_j\right|^2,
\end{align}
where $\approx$ are made in the limit of $|\alpha_k-\alpha_j|\gg1$ such that $f_{jk}\approx \delta_{jk}$. Then $\bar{\alpha}=0$ is equivalent to the weighted sum of all coherences to be zero.  $\bar{n}=|\bar{\xi}|$ limits the choice of the phases of $\alpha_j$ such that $\alpha_j^2$ have to be all in the same orientation. 

\begin{figure}[t]
\leavevmode\includegraphics[width = 0.8 \columnwidth]{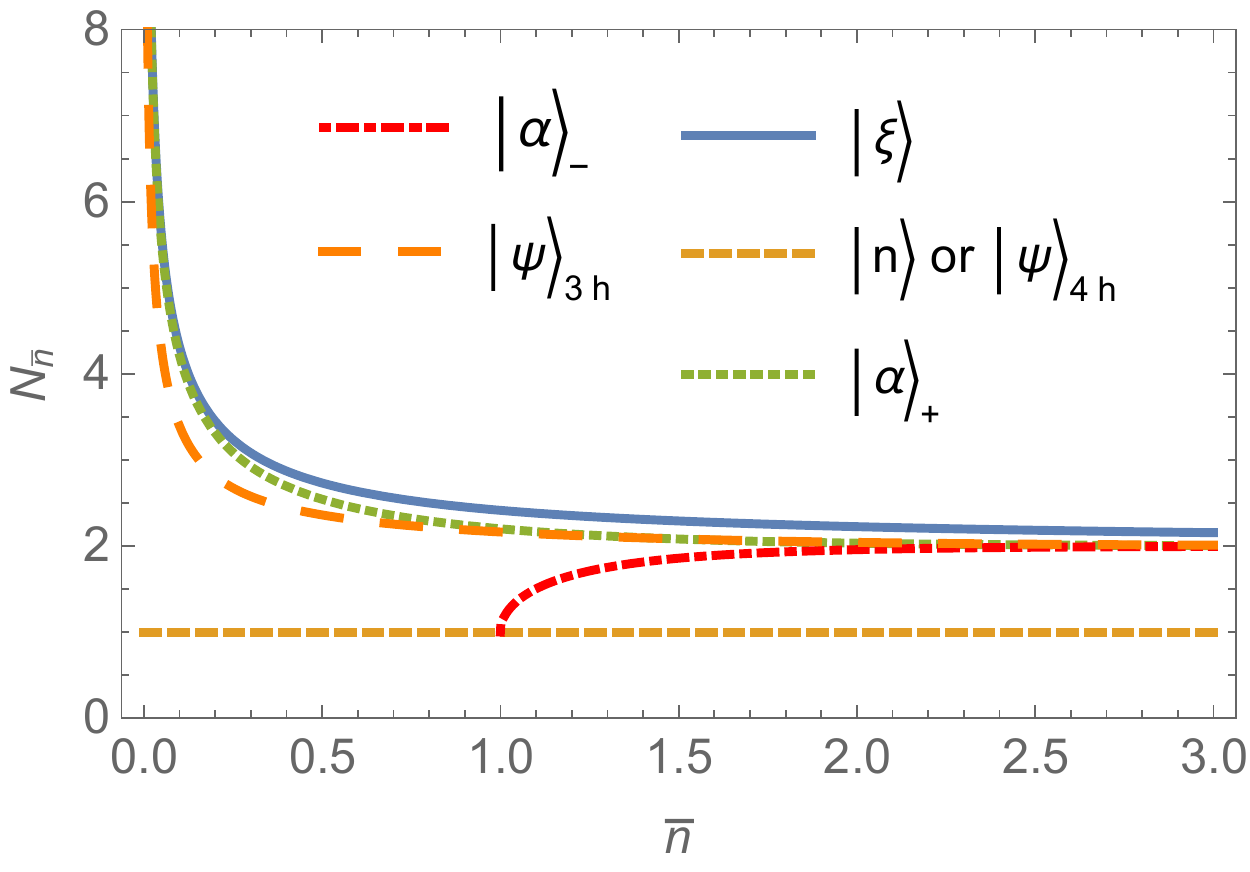}
\caption{Nonclassicality per unit energy $\mathcal{N}_{\bar{n}}$ for different pure states as a function of $\bar{n}$.} 
\label{fig:Npure} 
\end{figure} 

For concreteness, we list some of these states.
They can be even and odd cat states $\ket{\alpha}_{\pm}= N_{\pm}^{-1/2}\left(\ket{\alpha}\pm\ket{-\alpha}\right)$ \cite{Dodonov:1974aa} with $N_{\pm}=2\pm2 e^{-2|\alpha|^2}$ and their nonclassicalities per unit energy are given by $1+N_{\pm}/N_{\mp}$. Another example is a three-headed cat state $\ket{\psi}_{3h}=N_{3h}^{-1/2}\left(\ket{\alpha}+\ket{0}+\ket{-\alpha}\right)$ with $\mathcal{N}_{\bar{n}}\left(\ket{\psi}_{3h}\right)=1+(N_++2e^{-|\alpha|^2/2})/N_-$, where $N_{3h}=3+4e^{-|\alpha|^2/2}+2e^{-2|\alpha|^2}$. The values of $\mathcal{N}_{\bar{n}}$ of these states all approach to $2$ for $\bar{n}\gg1$ (Fig. \ref{fig:Npure}).

Our results show that there is a class of nonclassical states that are as useful as a squeezed vacuum for phase sensing in the MZI in the asymptotical limit, which are important for a number of experiments, such as gravitational wave detections using the LIGO  \cite{PhysRevLett.124.171101,PhysRevLett.124.171102, Aasi:2013aa}.

\subsubsection{Class 2: $1\le\lim_{\bar{n}\rightarrow \infty}\mathcal{N}_{\bar{n}}<2$}
The second class of nonclassical states are also useful for Heisenberg-limited sensing in the MZI. To find states that belong to this class, we consider two approaches: (i) adding a coherent displacement $\mathcal{D}(\alpha)$ \cite{SZ} onto the states in the first class; (ii) searching for the conditions that $\bar{\alpha}=0$ and $\bar{\xi}<\bar{n}$ according to Eq. \eqref{Npure}.

 \begin{figure}[t]
\leavevmode\includegraphics[width = 0.8 \columnwidth]{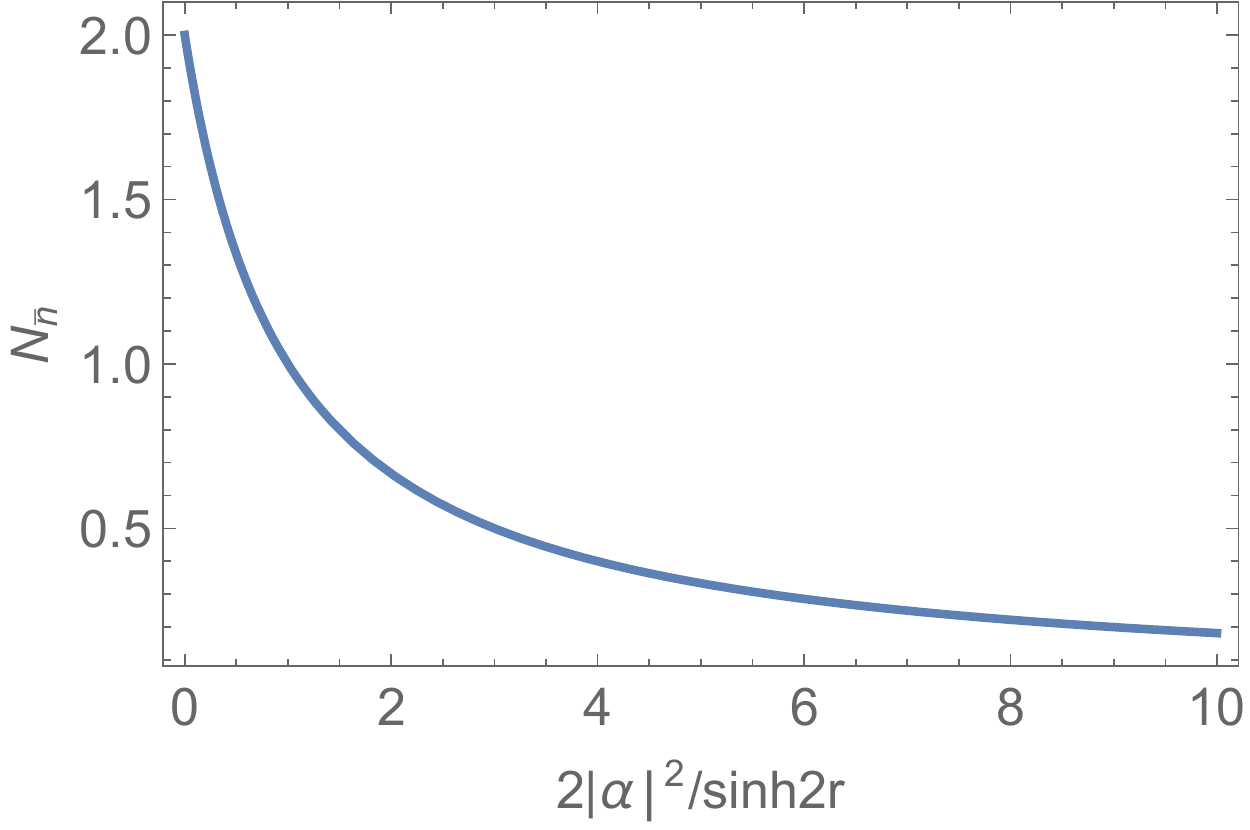}
\caption{Nonclassicality per unit energy of squeezed coherent states $\ket{\alpha,\xi}$ as a function of $|\alpha|^2$ normalized with respect to $\left(\sinh 2r\right)/2$ at $r=5$. For $2|\alpha|^2\lesssim\sinh 2r$, $\mathcal{N}_{\bar{n}}\gtrsim1$, indicating the ability for Heisenberg-limited sensing.} 
\label{fig:SC} 
\end{figure} 

The nonclassicality $\mathcal{N}$ is fixed by adding a coherent displacement since it is a classical operation \cite{Tan17}, while the nonclassicality per unit energy $\mathcal{N}_{\bar{n}}$ decreases. We give an example using the squeezed coherent states $\ket{\alpha,\xi}=\mathcal{D}(\alpha)\mathcal{S}(\xi)\ket{0}$. For a squeezed coherent state, we calculate its nonclassicality to be $\mathcal{N}\left(\ket{\alpha,\xi}\right)=\sinh^2r+\cosh r\sinh r$, where $\xi=re^{i\theta}$. Obviously, $\mathcal{N}\left(\ket{\alpha,\xi}\right)$ is independent of the coherent displacement. The nonclassicality per unit energy of the state is given by
\begin{align}
\mathcal{N}_{\bar{n}}\left(\ket{\alpha,\xi}\right)=\frac{\sinh^2r+\cosh r\sinh r}{|\alpha|^2+\sinh^2r}.
\end{align}
When we increase $|\alpha|$, the nonclassicality per unit energy will be reduced. For $0<|\alpha|^2\le\cosh r\sinh r$, we find $1\le\mathcal{N}_{\bar{n}}\left(\ket{\alpha,\xi}\right)<2$ for $\bar{n}\gg1$ (Fig. \ref{fig:SC}).

In the second approach, we consider a few examples that the conditions $\bar{\alpha}=0$ and $\bar{\xi}<\bar{n}$ are met. For example, Fock states have nonclassicalities simply given by $\mathcal{N}(\ket{n})=n$ according to Eq. \eqref{Npure}. They are an important class of nonclassical states which are also non-Gaussian \cite{PhysRevA.76.042327} and they can allow Heisenberg-limited phase sensing in a interferometer together with another input state, classical \cite{Xu:2012aa,PhysRevLett.110.163604} or nonclassical \cite{Holland:1993aa, GePRL2018}. A superposition of two Fock states \cite{Ryl17} $\ket{\psi(n)}=1/\sqrt{2}\left(\ket{0}+\ket{n}\right)$ can have nonclassicality $\mathcal{N}_{\bar{n}}(\ket{n})=1+\delta_{0,2}/\sqrt{2}$, which can also allow Heisenberg-limited sensitivity  \cite{Mccormick2018}. For another example, a four-headed cat state $\ket{\psi}_{4h}=N_{4h}^{-1/2}\left(\ket{\alpha}-\ket{i\alpha}+\ket{-\alpha}-\ket{-i\alpha}\right)$ with $N_{4h}=4-8e^{-|\alpha|^2}\cos|\alpha|^2+4e^{-2|\alpha|^2}$, which is also useful for quantum error correction \cite{PhysRevLett.119.030502}, has the same nonclassicality as a Fock state for the same amount of energy, i.e., $\mathcal{N}_{\bar{n}}\left(\ket{\psi}_{4h}\right)=1$. 
The nonclassicalities per unit energy of these states are shown in Fig. \ref{fig:Npure} as a function of $\bar{n}$. 

\begin{widetext}
\renewcommand{\arraystretch}{1.4}
\setlength{\tabcolsep}{7pt}

\begin{table}
\caption{Nonclassicality with Single-photon addition}
\begin{ruledtabular}
\begin{tabular}{c c c c c}
\label{tb}
& $\ket{\alpha}$    &  $\ket{\xi}$ & $\ket{\alpha}_{\pm}$ \\
\hline
$\mathcal{N}$ & $0$  & $\frac{1}{2}(e^{2r}-1)$  & $|\alpha|^2\left(1+N_{\mp}/N_{\pm}\right)$\\[0.75ex] 
$\mathlarger{\mathcal{N}_{\bar{n}}}$ & $0$  & $1+\sqrt{1+\frac{1}{\bar{n}}}$ & $1+N_{\pm}/N_{\mp}$\\[1.25ex]
$\mathcal{N}[\hat{a}^{\dagger}]$ & $\mathlarger{\frac{1}{1+|\alpha|^2}}$  & $\frac{1}{2}(3e^{2r}-1)$  & $ \mathlarger{\frac{|\alpha|^2\left(|\alpha|^2+3\right)\left(1+\frac{N_{\mp}}{N_{\pm}}\right)+1}{|\alpha|^2N_{\mp}/N_{\pm}+1}}$\\[3ex]
$\mathlarger{\mathcal{N}[\hat{a}^{\dagger}]_{\bar{n}}}$ & $\mathlarger{\frac{1}{\left(1+|\alpha|^2\right)^2}}$ & $1+\sqrt{1+\frac{1}{\bar{n}}-\frac{2}{\bar{n}^2}}$ & $\mathlarger{\frac{|\alpha|^2\left(|\alpha|^2+3\right)\left(1+\frac{N_{\mp}}{N_{\pm}}\right)+1}{|\alpha|^2\left(|\alpha|^2+3N_{\mp}/N_{\pm}\right)+1}}$
\end{tabular}
\end{ruledtabular}
\label{tab}
\end{table}
\end{widetext}

\subsubsection{Class 3: $0<\lim_{\bar{n}\rightarrow \infty}\mathcal{N}_{\bar{n}}<1$}
This class of states are less nonclassical than the previous two classes. Nevertheless, they can provide sensitivity beyond the SQL in the sensing tasks. States that fall in category must have nonzero coherence, i.e., $\bar{\alpha}\ne0$. For example, they can be coherent squeezed states with $|\alpha|^2>\cosh r\sinh r$, $0<\mathcal{N}_{\bar{n}}\left(\ket{\alpha,\xi}\right)<1$ (see Fig. \ref{fig:SC}). 

In summary, we have categorized the nonclassical pure states into three classes using the nonclassicality measure per unit energy. This provides a useful reference for comparing nonclassicality for any pure states in terms of quantum metrology.

\subsubsection{Photon-added states}
Now we study the nonclassicality of a state after single-photon addition \cite{PhysRevA.43.492, Zavatta660,PhysRevA.82.063833, Ryl17,PhysRevA.91.022317}, which can be achieved via conditioned nonlinear interaction with a single two-level atom \cite{PhysRevA.43.492} or a parametric crystal \cite{Zavatta660,PhysRevA.82.063833}. Single-photon addition has been demonstrated to generate nonclassical features using classical states \cite{Zavatta660,PhysRevA.82.063833}. Here we examine quantitatively the nonclassicalities of a quantum state before and after single-photon addition (indicated by $\mathcal{N}[\hat{a}^{\dagger}]$ and $\mathlarger{\mathcal{N}[\hat{a}^{\dagger}]_{\bar{n}}}$ in Table \ref{tb}).

 \begin{figure}[t]
\leavevmode\includegraphics[width = 0.8 \columnwidth]{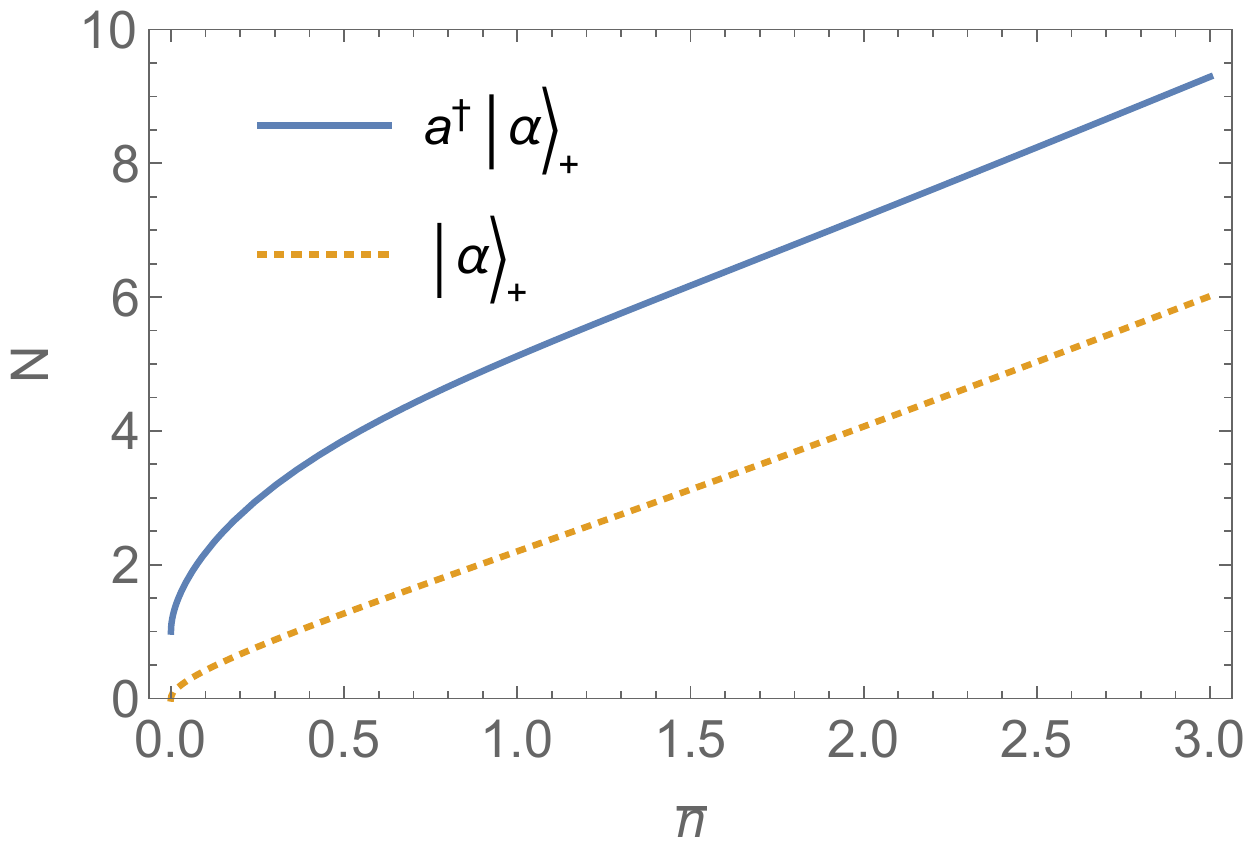}
\caption{Nonclassicality of an even cat state $\ket{\alpha}_+$ and that after a single-photon addition as a function of $\bar{n}$.} 
\label{fig:Cat-A} 
\end{figure} 

 \begin{figure}[t]
\leavevmode\includegraphics[width = 0.8 \columnwidth]{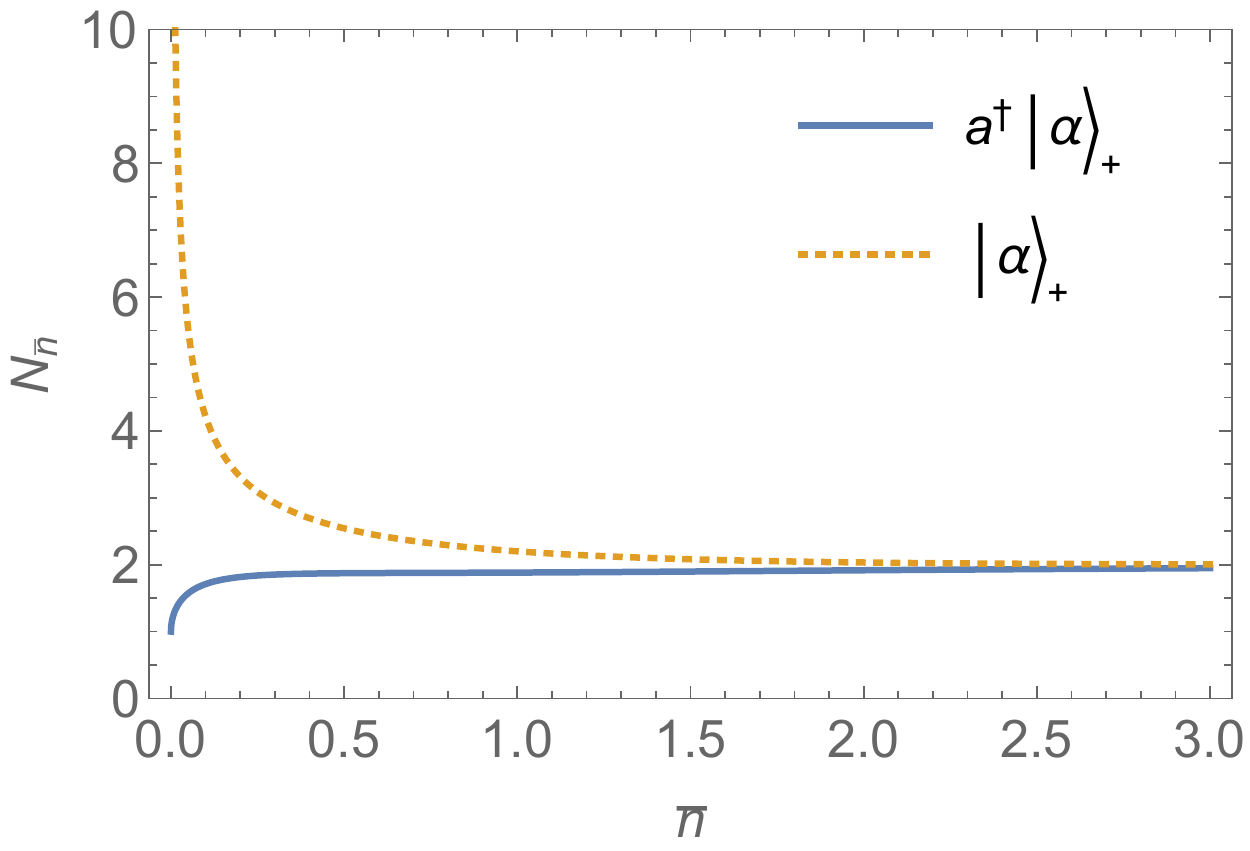}
\caption{Nonclassicality per unit energy of an even cat state $\ket{\alpha}_+$ and that after a single-photon addition as a function of $\bar{n}$.} 
\label{fig:Cat-APer} 
\end{figure} 

We observe that the nonclassicality is increased for all the states we studied. For example, single-photon added coherent state $\ket{1,\alpha}\equiv\frac{\hat{a}^{\dagger}}{\sqrt{1+|\alpha|^2}}\ket{\alpha}$  \cite{PhysRevA.82.063833, Zavatta660} has a nonclassicality of $1/(1+|\alpha|^2)$ comparing with zero of a coherent state. We note that the nonclassicality of $\ket{1,\alpha}$ is smaller than that of a coherent-displaced single-photon state $\ket{\alpha,1}=\mathcal{D}(\alpha)\ket{1}$. The reason is that the former is a superposition of a coherent state (a classical state) and $\ket{\alpha,1}$, i.e. $\ket{1,\alpha}=\left(\alpha^{\ast}\ket{\alpha}+\ket{\alpha,1}\right)/\sqrt{1+|\alpha|^2}$, so that its nonclassicality is between $0$ and $1$. 
Interestingly, for initial nonclassical states in Class 1 ($\lim_{\bar{n}\rightarrow \infty}\mathcal{N}_{\bar{n}}=2$), the nonclassicality enhancement, $\mathcal{N}[\hat{a}^{\dagger}]- \mathcal{N}$ , after single-photon addition can be much greater than $1$. For example, the nonclassicality of single-photon added squeezed vacuum is almost three time larger than its value before the operation (Table \ref{tb}). Similarly, the nonclassicality of even cat states is increased to twice its value when $\bar{n}=2$ as shown in Fig. \ref{fig:Cat-A}. This result suggests a potential protocol for preparing a stronger nonclassical state via single-photon addition.

We also observe that the nonclassicality per unit energy, $\mathcal{N}_{\bar{n}}$, after single-photon addition is between the value of the single-photon state and that of the state before the operation. For example, $0<\mathcal{N}_{\bar{n}}(\ket{1,\alpha})<1$, and $1<\mathcal{N}_{\bar{n}}(\ket{1,\xi})<1+\sqrt{1++\frac{1}{\bar{n}}}$. For Fock states, the nonclassicality per unity energy is the same before and after the single-photon addition. For cat states, we provide the derivations in Table \ref{tb}. However, the analytical results are difficult to see this feature, so we plot the even cat state before and after the single-photon addition as an example in Fig. \ref{fig:Cat-APer}. The nonclassicality per unit energy decreases after single-photon addition for the even cat state although the difference is marginal for $\bar{n}\gg1$. We have checked this statement for a single-photon added odd cat state and we conjecture it to be true for any pure states.


\subsection{Nonclassicality for mixed states\label{sec3b}}
Although pure states are ideal for various applications, mixed states are inevitable in practical situations due to their coupling to the environment. According to Eq. \eqref{eq:n-mixed}, evaluating single-mode nonclassicality for mixed states is a nontrivial task as one needs to find the minimum value of the expression among all possible decomposition of $\hat\rho$, while the number of decompositions can be infinite in principle. 

To evaluate the ORT measure, we analyze the structure of a special class of mixed states that can be written in diagonal Fock basis. We study some properties of these states and derive the nonclassicalities of some nontrivial examples. In the situation when directly evaluating the nonclassicality measure is challenge, for example a mixed state with an infinite dimensions, we can lower bound $\mathcal{N}\left(\hat\rho\right)$ using the metrological power $\mathcal{W}\left(\hat\rho\right)$ from QFI via Eq. \eqref{eq:ineq1}. We provide an example using a single-photon added thermal state. 

We consider those states that can be written in diagonal Fock basis, i.e. $\hat\rho=\sum_{i=0}^L p_i\ket{i}\bra{i}$, where $\sum_i^L p_i=1$ and $p_i\ge0$. Obviously, $\{p_i,\ket{i}\}$ is one possible decomposition. Therefore, any decomposition of $\hat\rho$ can be given by
\begin{align}
\hat{\rho}=\sum_{j=0}^Kq_j\ket{\phi_j}\bra{\phi_j},
\end{align}
where $\ket{\phi_j}=1/\sqrt{q_j}\sum_{i=0}^LU_{ij}\sqrt{p_i}\ket{i}$, $q_j=\sum_{i=0}^LU_{ij}U_{ij}^{\ast}p_i$, and $K\ge L$. The matrix $U$ is an isometry such that $\sum_{j=0}^KU_{ij}U_{i^{\prime}j}^{\ast}=\delta_{i,i^{\prime}}$. 

 \begin{figure}[t]
\leavevmode\includegraphics[width = 0.8 \columnwidth]{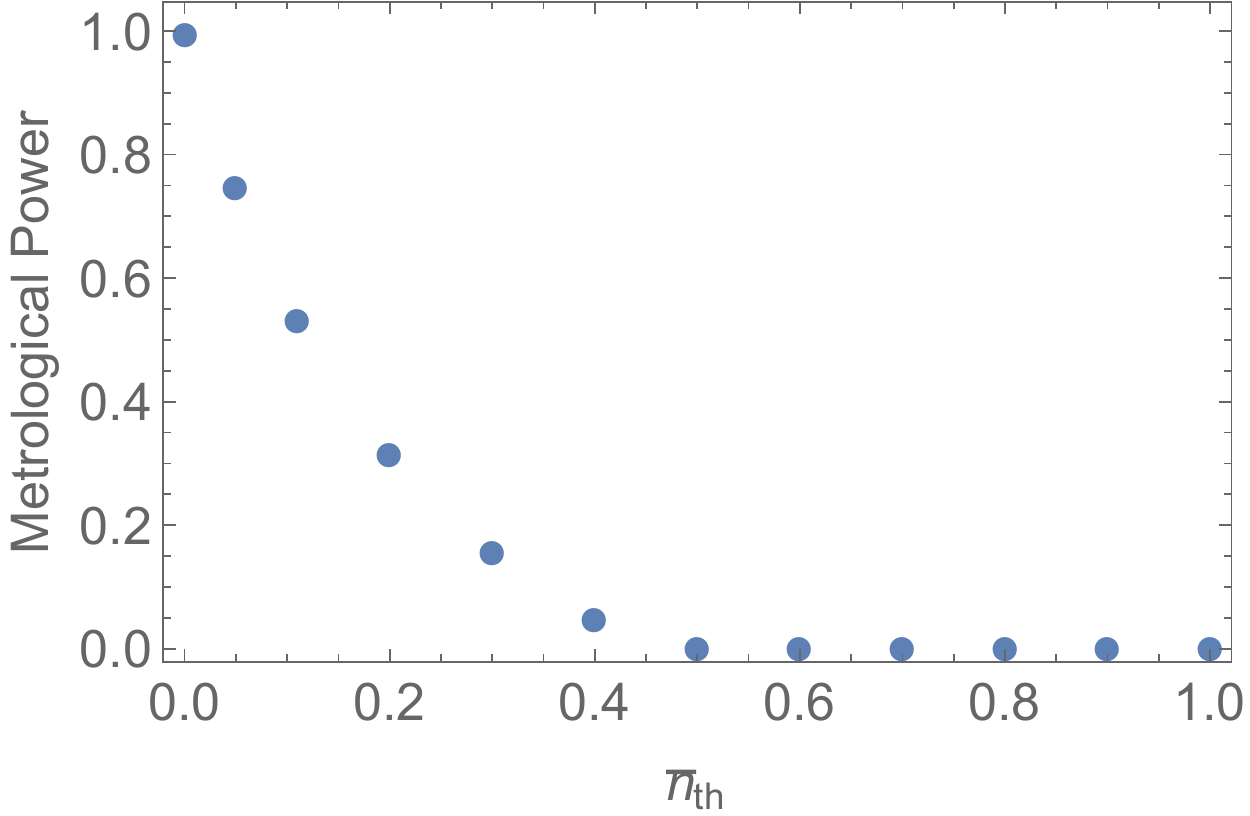}
\caption{The metrological power $\mathcal{W}$ of single-photon-added thermal state $\hat\rho_{\text{th},1}$ as a function of mean thermal photon number $\bar{n}_{\text{th}}$. The metrological power transits from positive values to zero as  $\bar{n}_{\text{th}}$ increases.} 
\label{fig:Wnbar} 
\end{figure} 

For any decomposition, we have $\bar{n}_j=1/q_j\sum_{i=0}^LU_{ij}U_{ij}^{\ast}p_ii$, $\bar{\alpha}_j=1/q_j\sum_{i=0}^LU_{ij}U_{i-1j}^{\ast}\sqrt{p_ip_{i-1}i}$, and $\bar{\xi}_j=1/q_j\sum_{i=0}^LU_{ij}U_{i-2j}^{\ast}\sqrt{p_ip_{i-2}i(i-1)}$. Interestingly, we find that for any decomposition set $\{q_j,\ket{\phi_j}\}$, the following properties hold
\begin{align}
\sum_{j=0}^Kq_j\bar{n}_j=\braket{\hat{a}^{\dagger}\hat{a}},\qquad  \sum_{j=0}^Kq_j\bar{\alpha}_j=0, \qquad \sum_{j=0}^Kq_j\bar{\xi}_j=0,
\end{align}
where $\braket{\hat{a}^{\dagger}\hat{a}}=\sum_{i=0}^Lp_ii$. According to its definition Eq. \eqref{eq:n-mixed}, we obtain that
\begin{align}
\mathcal{N}\left(\hat\rho\right)=\braket{\hat{a}^{\dagger}\hat{a}}-\max_{\{q_j,\ket{\phi_j}\}}\biggl\{\sum_{j=0}^K q_j|\bar{\alpha}_j|^2-\biggl|\sum_{j=0}^K q_j\bar{\alpha}_j^2\biggr|\biggr\}\le\braket{\hat{a}^{\dagger}\hat{a}}.
\end{align}
The inequality can be achieved when $\bar{\alpha}_j=0$ and some trivial solutions are the state $\hat\rho$ with $p_ip_{i-1}=0$ for any $i$. For example, $\hat\rho=\sum_{i=0}^L p_{2i}\ket{2i}\bra{2i}$, which can be a completely phase-damped squeezed vacuum state with $p_{2i}=\left(\cosh r\right)^{-1}\frac{\left(2i\right)!}{\left(i!\right)^2}\left(\frac{1}{2}\tanh r\right)^{2i}$. Another example is the state $\hat\rho=p\ket{0}\bra{0}+(1-p)\ket{n}\bra{n}$ ($n>2$), whose nonclassicality equals to its pure state counterpart $1/\sqrt{p}\ket{0}+1/\sqrt{1-p}\ket{n}$. This suggests that certain the nonclassicality of Fock state superpositions may not be affected by a phase-damped environment \cite{PhysRevA.62.053807}.

On the other hand, the nonclassicality is lowered bounded via the relation
\begin{align}
\mathcal{N}\left(\hat\rho\right)&\ge\braket{\hat{a}^{\dagger}\hat{a}}-\max_{\{q_j,\ket{\phi_j}\}}\sum_{j=0}^K q_j|\bar{\alpha}_j|^2\end{align}
when $\sum_{j=0}^K q_j\bar{\alpha}_j^2=0$. Now we consider some examples to achieve this lower bound. If the state consists of two nearest Fock basis, e.g,
\begin{align}
\hat\rho_{2F}=(1-p)\ket{n+1}\bra{n+1}+p\ket{n}\bra{n},
\end{align}
it can be decomposed via $\ket{\phi_j}=1/\sqrt{2q_j}\left(\sqrt{1-p}e^{i\varphi_j}\cos\theta_j\ket{n+1}+\sqrt{p}\sin\theta_j \ket{n}\right)$ ($j=0,1,2,3$) and choosing $\varphi_0=\varphi_1=\varphi_2-\pi/2=\varphi_3-\pi/2$ and $\theta_0=\theta_1-\pi/2=\theta_2=\theta_3-\pi/2$. We find 
\begin{align}
&\max_{\{q_j,\ket{\phi_j}\}}\sum_{j=0}^K q_j|\bar{\alpha}_j|^2\nonumber\\
=&\max_\theta \left\{\frac{(n+1)p(1-p)\sin^2\theta\cos^2\theta}{\left[(1-p)\cos^2\theta+p\sin^2\theta\right]\left[p\cos^2\theta+(1-p)\sin^2\theta\right]}\right\}\nonumber\\
=&(n+1)p(1-p).
\end{align}
So the nonclassicality of the state is $\mathcal{N}\left(\hat\rho_{2F}\right)=(n+1)(1-p)^2+np$.

For a mixed state with components $L\ge3$, the analytical optimization can be more difficult. Instead, we use the operational relation Eq. \eqref{eq:ineq1} to lower-bound the nonclassicality $\mathcal{N}$ of $\hat\rho=\sum_{i=0}^L p_i\ket{i}\bra{i}$, which is given by
\begin{align}
\mathcal{N}\left(\hat\rho\right)\ge\mathcal{W}\left(\hat\rho\right)=\max\left\{\sum_{i=1}^{L}\frac{ip_i\left(p_i-p_{i-1}\right)}{p_i+p_{i-1}},0\right\}.
\end{align}


\begin{widetext}
\renewcommand{\arraystretch}{1.4}
\setlength{\tabcolsep}{7pt}
 
\begin{table}
\caption{Comparison of Nonclassicality Measures for a pure state $\ket{\psi}=\sum_{k=1}^Lc_k\ket{\alpha_k}$}
\begin{ruledtabular}
\begin{tabular}{c c c c c c}
 &Nonclassical depth $\tau$ \cite{Lee:91}    &  Degree of Nonclassicality \cite{GehrkePRA12} & RT measure $Q$ \cite{YadinPRX18, Kwon19} & ORT measure $\mathcal{N}$ \cite{ge2019operational} \\
\hline
                        & Convolution parameter \\[-0.75ex]
Definition        & for a positive-definite & $E_\text{Ncl}=1-e^{-L+1}$  & $2\left(\bar{n}-|\alpha|^2\right)$ & $\bar{n}-|\bar{\alpha}|^2+\left|\bar{\xi}-\bar{\alpha}^2\right|$  \\[-0.75ex] 
                      &quasi-prob. distribution  \\[-0.25ex]
Operationality & unknown& unknown & Pure states only &  An arbitrary state\\[0.25ex]
Maximum value & $1$  & $1$ & $2\bar{n}$ & $\bar{n}+\sqrt{\bar{n}(\bar{n}+1)}$\\[0.25ex]
Most nonclassical states & Non-Gaussian states  & States with $r=\infty$  & $\ket{n},\ \ket{\xi}, \ \ket{\alpha}_{\pm}$, etc. & $\ket{\xi}$\\[0ex]
Distinguishability of $\ket{n}$ & No& Yes & Yes &  Yes\\
Distinguishability of $\ket{\xi}$ & Yes& No & Yes &  Yes\\
\begin{tabular}{c}
Distinguishability \\[-0.5ex]
 between $\ket{n}$ and $\ket{\xi}$ 
 \end{tabular}  & Yes& Yes & No &  Yes
\end{tabular}
\end{ruledtabular}
\label{tab2}
\end{table}
\end{widetext}

\noindent We observe from the above expression that $\mathcal{W}$ can be zero even for a nonclassical mixed state, meaning there is no metrological advantage for sensing tasks using certain nonclassical states, which falls into the region between the red and the green ovals in Fig. \ref{fig:state}. An example we give is a single-photon added thermal state $\hat\rho_{\text{th},1}$ with $p_i=\frac{i}{\bar {n}^2}\left(\frac{\bar{n}}{\bar{n}+1}\right)^{i+1}$, which is shown to be nonclassical for any positive value of $\bar n_{\text{th}}$ \cite{PhysRevA.75.052106, PhysRevA.83.032116}. However, $\mathcal{W}=0$ for $\bar{n}_{\text{th}}>0.4567$ (see Fig. \ref{fig:Wnbar}).

\section{Discussion\label{sec4}}
We summarize some interesting properties of the ORT measure of nonclassicality in comparison with those of some important existing measures in Table \ref{tab2}. We give an example by discussing the nonclassicalities of Fock states and squeezed vacuum states using these measures.

According to the nonclassicality depth \cite{Lee:91}, Fock states have the same nonclassical depth of $1$, which is always greater than any nonclassical Gaussian states, e.g., squeezed vacuum states, of which the nonclassical depth is between $0$ and $1/2$. The nonclassicality depth concludes Fock states are more nonclassical than squeezed vacuum states regardless of the squeezing strength. The conclusion is exactly the opposite using the definition via the Schmidt rank \cite{GehrkePRA12}. In that definition, any squeezed vacuum state has the maximum nonclassicality independent of its squeezing strength $r$, while those of Fock states $\ket{n}$ increase with $n$ and are bounded by the maximum value. Using the RT measure measure $Q$ \cite{YadinPRX18, Kwon19}, the nonclassicality of Fock states and squeezed vacuum states are equal for a fixed energy and it grows with the energy.

Using the ORT measure, we draw some different conclusions. First, the ORT nonclassicality measure can be used to compare non-Gaussian states (Fock states) with that of Gaussian states (squeezed vacuum states) from the operational meaning. Second, the measure for both Fock states and squeezed vacuum states increases with its energy, but it increases at different rates such that the squeezed vacuum states are more nonclassical for a fixed energy. One implication is that the nonclassicalities of these states can be the same when a Fock state has more energy than that of a squeezed vacuum.

\section{Conclusion\label{sec5}}
We have evaluated extensively the nonclassicality of single-mode quantum states using the ORT measure, which satisfies the minimum requirements of the resource theory of nonclassicality and directly relates to the ability for quantum-enhanced sensing tasks. 

For pure states, the ORT measure quantifies the ability of quadrature sensing and phase sensing in the balanced interferometer and we have investigated the measure for different classes of states. In particular, we have discovered a class of nonclassical states that can attain the maximum sensing ability in the asymptotic limit of large energy. We have also found that single-photon additions can greatly improve nonclassicality of a quantum state, hence its ability for quantum sensing. For mixed states, the measure is difficult to evaluate as one needs to find the convex roof expression over an infinite possibilities of decompositions of a quantum state $\hat\rho$. Nevertheless, we have studied some nontrivial examples by analytically deriving the measure, which provides some idea of evaluating nonclassicality for mixed states. 

Our work has taken a step further in quantitatively understanding nonclassicality, which will be important for fundamental research in quantum optics and practical applications in quantum technologies. In particular, our results on evaluating single-mode nonclassicality will have useful applications in quantum sensing tasks.

\begin{acknowledgments}
This research is supported by a grant from King Abdulaziz City for Science and Technology (KACST).
\end{acknowledgments}

\bibliography{Noncl-Distri}

\end{document}